\documentstyle[preprint,aps,epsf,floats]{revtex}
\begin{document}

\def\lqcd{\Lambda_{\rm QCD}}
\def\DKsln{D\to K^* \bar\ell \nu}
\def\Drln{D\to \rho \bar\ell \nu}
\def\BKsg{\bar B\to K^* \gamma}
\def\Brg{\bar B\to \rho \gamma}
\def\BKsll{\bar B\to K^* \ell \bar\ell}
\def\Brln{\bar B\to \rho \ell \bar\nu}

\preprint{\tighten \vbox{\hbox{FERMILAB-Pub-99/142-T} \hbox{CALT-68-2224}
  \hbox{hep-ph/9905277} \hbox{} }}

\title{$\BKsg$ from $\DKsln$ }
 
\author{Zoltan Ligeti$\,^a$ and Mark B.\ Wise$\,^b$ }

\address{
  $^a$Theory Group, Fermilab, P.O.\ Box 500, Batavia, IL 60510 \\[-8pt]
  $^b$California Institute of Technology, Pasadena, CA 91125  }

\maketitle

{\tighten
\begin{abstract}%
The $\BKsg$ branching fraction is predicted using heavy quark spin symmetry at
large recoil to relate the tensor and (axial-)vector form factors, using heavy
quark flavor symmetry to relate the $B$ decay form factors to the measured
$\DKsln$ form factors, and extrapolating the semileptonic $B$ decay form
factors to large recoil assuming nearest pole dominance.  This prediction
agrees with data surprisingly well, and we comment on its implications for the
extraction of $|V_{ub}|$ from $\Brln$.

\end{abstract}
}%end tighten

\newpage

The next generation of $B$ decay experiments will test the
Cabibbo--Kobayashi--Maskawa (CKM) picture of quark mixing and $CP$ violation
with high precision.  The basic approach is to determine the sides and angles
of the unitarity triangle, and then check for the consistency of these results.
A precise and model independent determination of the magnitude of the $b \to u$
CKM matrix element, $|V_{ub}|$, is particularly important.  It is one of the
least precisely known elements of the CKM matrix.  At the present time the
uncertainty of the standard model expectation for $\sin(2\beta)$, the $CP$
asymmetry in $B\to J/\psi K_S$, depends strongly on the uncertainty of
$|V_{ub}|$.

Currently, most determinations of $|V_{ub}|$ rely on phenomenological
models~\cite{exptvub}.  The more promising model independent approaches for
the future include studying the hadronic invariant mass distribution in
inclusive semileptonic $\bar B\to X_u e\bar\nu$ decay~\cite{FLW}, measuring the
inclusive $\bar B\to X_{u\bar cd}$ nonleptonic decay rate~\cite{FP}, and
comparing the exclusive $\Brln$ and $\bar B\to \pi \ell \bar\nu$ decay rates in
the large $q^2$ region with lattice results~\cite{lattice} or predictions based
on heavy quark symmetry and chiral symmetry~\cite{IsWi,lwold,lsw}.  A major
uncertainty in the latter method is the size of the symmetry breaking
corrections.  Another question for this approach is whether the $\DKsln$ (or
$\Drln$) form factors can be extrapolated to cover a larger fraction of the
$\Brln$ phase space.

In this paper some of these ingredients are tested by comparing the measured
$\BKsg$ branching fraction with a prediction relying on $b$ quark spin symmetry
at large recoil to relate the tensor and (axial-)vector form factors, heavy
quark flavor symmetry to relate the $B$ decay form factors to the measured
$\DKsln$ form factors, and an extrapolation of the semileptonic $B$ decay form
factors assuming nearest pole dominance.  We denote by a superscript $(H\to V)$
the form factors relevant for transitions between a pseudoscalar meson $H$
containing a heavy quark, $Q$, and a member of the lowest lying multiplet of
vector mesons, $V$.  We view the form factors as functions of the dimensionless
variable $y = v\cdot v'$, where $p=m_H\,v$, $p'=m_V\,v'$, and $q^2 = (p-p')^2 =
m_H^2 +m_V^2 -2m_H\,m_V\,y$.  (Note that even though we are using the variable
$v\cdot v'$, we are not treating the quarks in $V$ as heavy.)  An approach with
some similarities to the one presented here can be found in Ref.~\cite{other}. 
This decay has also been considered in Refs.~\cite{other1,other2}.

The $\BKsg$ transition arises from a matrix element of the effective 
Hamiltonian,
\begin{equation}
H_{\rm eff} = -{4G_F \over \sqrt2}\, V_{ts}^*\,V_{tb}\,
  \sum_{i=1}^8 C_i(\mu)\, O_i(\mu) \,,
\end{equation}
where $G_F$ is the Fermi constant, and $C_i (\mu)$ are Wilson coefficients
evaluated at a subtraction point $\mu$.  The $\BKsg$ matrix element of $H_{\rm
eff}$ is thought to be dominated by the operator
\begin{equation}
O_7 = {e \over 16\pi^2}\, \overline{m}_b\, 
  \bar s_L\,\sigma^{\mu\nu} F_{\mu\nu}\,b_R \,,
\end{equation}
where $e$ is the electromagnetic coupling, $\overline{m}_b$ is the
$\overline{\rm MS}$ $b$ quark mass, and $F_{\mu\nu}$ is the electromagnetic
field strength tensor.  $O_1-O_6$ are four-quark operators and $O_8$ involves
the gluon field strength tensor.

The $\BKsg$ matrix element of $O_7$ can be expressed in terms of hadronic form
factors, $g_\pm$ and $h$, defined by
\begin{eqnarray}
\langle V(p',\epsilon) |\,\bar q\,\sigma_{\mu\nu}\, Q\,| H(p)\rangle
&=& g_+^{(H\to V)}\, \varepsilon_{\mu\nu\lambda\sigma}\, \epsilon^{*\lambda}\,
  (p+p')^\sigma + g_-^{(H\to V)}\, \varepsilon_{\mu\nu\lambda\sigma}\,
  \epsilon^{*\lambda}\, (p-p')^\sigma \nonumber\\*
&+& h^{(H\to V)}\, \varepsilon_{\mu\nu\lambda\sigma}\, (p+p')^\lambda\, 
  (p-p')^\sigma\, (\epsilon^*\cdot p) \,, \nonumber\\
\langle V(p',\epsilon) |\,\bar q\,\sigma_{\mu\nu}\gamma_5\, Q\,| H(p)\rangle
&=& i\,g_+^{(H\to V)}\,[\epsilon^*_\nu\,(p+p')_\mu-\epsilon^*_\mu\,(p+p')_\nu]
  \nonumber\\*
&+& i\,g_-^{(H\to V)}\,[\epsilon^*_\nu\,(p-p')_\mu-\epsilon^*_\mu\,(p-p')_\nu]
  \nonumber\\*
&+& i\,h^{(H\to V)}\, [(p+p')_\nu\,(p-p')_\mu-(p+p')_\mu\,(p-p')_\nu]\,
  (\epsilon^*\cdot p) \,.
\end{eqnarray}
The second relation follows from the first one using the identity
$\sigma^{\mu\nu} = \frac i2\, \varepsilon^{\mu\nu\alpha\beta}
\sigma_{\alpha\beta} \gamma_5$.  We use the convention $\varepsilon^{0123} =
-\varepsilon_{0123}=1$.  The $\BKsg$ decay rate is then given by
\begin{equation}\label{Rrate}
\Gamma(\BKsg) = {G_F^2\, \alpha\, |V_{ts}^*V_{tb}|^2\over 32\,\pi^4}\,
  {\overline m}_b^2\, m_B^3\, \bigg( 1 - {m_{K^*}^2\over m_B^2} \bigg)^3\,
  |C_7|^2\, \Big| g_+^{(B\to K^*)}(y_0) \Big|^2 \,,
\end{equation}
where $y_0 = (m_B^2 + m_{K^*}^2) / (2 m_B m_{K^*}) = 3.05$.

In semileptonic decays such as $\DKsln$ or $\Brln$ another set of form factors 
occur, $g$, $f$, and $a_\pm$, defined by
\begin{eqnarray}\label{ffdef}
\langle V(p',\epsilon) |\,\bar q\,\gamma_\mu\, Q\,| H(p)\rangle
&=& i\,g^{(H\to V)}\, \varepsilon_{\mu\nu\lambda\sigma}\, \epsilon^{*\nu}\,
  (p+p')^\lambda\, (p-p')^\sigma \,, \\*
\langle V(p',\epsilon) |\,\bar q\,\gamma_\mu\gamma_5\, Q\,| H(p)\rangle
&=& f^{(H\to V)}\,\epsilon^*_\mu 
  + a_+^{(H\to V)}\,(\epsilon^*\cdot p)\,(p+p')_\mu 
  + a_-^{(H\to V)}\,(\epsilon^*\cdot p)\,(p-p')_\mu \nonumber\,.
\end{eqnarray}
The experimental values for the $\DKsln$ form factors assuming nearest pole 
dominance for the $q^2$ dependences are \cite{E791b}
\begin{eqnarray}\label{ffexp}
f^{(D\to K^*)}(y) &=& {(1.9\pm0.1)\,{\rm GeV}\over 1+0.63\,(y-1)}\,, 
  \nonumber\\
a_+^{(D\to K^*)}(y) &=& -{(0.18\pm0.03)\,{\rm GeV}^{-1}\over 1+0.63\,(y-1)}\,, 
  \nonumber\\
g^{(D\to K^*)}(y) &=& -{(0.49\pm0.04)\,{\rm GeV}^{-1}\over 1+0.96\,(y-1)}\,.
\end{eqnarray}
The shapes of these form factors are beginning to be probed experimentally and
the pole form is consistent with data~\cite{E791b}.  The form factor $a_-$ is
not measured because its contribution to the $\DKsln$ decay amplitude is
suppressed by the lepton mass.  The minimal value of $y$ is unity
(corresponding to the zero recoil point) and the maximum value of $y$ is
$(m_D^2+m_{K^*}^2) / (2m_D\,m_{K^*}) \simeq 1.3$ (corresponding to $q^2=0$). 
In comparison, the allowed kinematic region for $\Brln$ is $1 < y < 3.5$.

A prediction for the $\BKsg$ decay rate can be made using heavy quark
spin symmetry, which implies relations between the tensor and (axial-)vector
form factors in the $m_b\to \infty$ limit~\cite{IsWi,lwold}
\begin{eqnarray}\label{spinrel}
g_+^{(B\to K^*)} + g_-^{(B\to K^*)} &=& {f^{(B\to K^*)} + 
  2 g^{(B\to K^*)}\,m_B\,m_{K^*}\,y \over m_B}\,,  \nonumber\\*
g_+^{(B\to K^*)} - g_-^{(B\to K^*)} &=& -2 m_B\, g^{(B\to K^*)} \,, \\*[4pt]
h^{(B\to K^*)} &=& {a_+^{(B\to K^*)} - a_-^{(B\to K^*)} - 
  2 g^{(B\to K^*)}\over 2\,m_B} \,, \nonumber
\end{eqnarray}
and therefore,
\begin{equation}\label{spinrel2}
g_+^{(B\to K^*)} = - g^{(B\to K^*)}\, (m_B - m_{K^*} y) 
  + f^{(B\to K^*)} / (2m_B) \,.
\end{equation}
We use heavy quark symmetry again to obtain $g^{(B\to K^*)}$ and 
$f^{(B\to K^*)}$ from the measured $\DKsln$ form factors given in 
Eq.~(\ref{ffexp})~\cite{IsWi}
\begin{eqnarray}\label{BDrel}
f^{(B\to K^*)}(y) &=& \left({m_B\over m_D}\right)^{1/2}\,
  f^{(D\to K^*)}(y)\,, \nonumber\\*
g^{(B\to K^*)}(y) &=& \left({m_D\over m_B}\right)^{1/2}\,
  g^{(D\to K^*)}(y)\,.
\end{eqnarray}
For $y$ not too large, Eq.~(\ref{spinrel}) has order $1/m_b$ corrections,
whereas Eq.~(\ref{BDrel}) receives both order $1/m_b$ and $1/m_c$ corrections. 

Model dependence in our prediction of $\Gamma(\BKsg)$ arises from the use of
$b$ quark spin symmetry at large recoil and due to the fact that the $B$ decay
form factors are extrapolated beyond $y=1.3$.  In Ref.~\cite{largew} it was
argued that the heavy quark spin symmetry relations in Eq.~(\ref{spinrel})
should hold over the entire phase space without unusually large corrections. 
To extrapolate $f^{(B\to K^*)}$ and $g^{(B\to K^*)}$ to values of $y>1.3$ we
assume the pole form, i.e., we simply use Eqs.~(\ref{ffexp}) and (\ref{BDrel})
evaluated at $y_0 = 3.05$.\footnote{The $y$-dependence of the nearest pole
dominated form factors for $B$ decay are expected to be almost the same as for
$D$ decay, so we continue to use Eq.~(\ref{ffexp}) for $y>1.3$.  For example,
with $m_{B_s^*}=5.42\,$GeV the ``slope" of $g^{(B\to K^*)}$ is 0.94 (instead of
0.96), and with $m_{B_s^{**}}=5.87\,$GeV the ``slope" of the axial form factors
are 0.62 (instead of 0.63).}  Although this is not a controlled approximation,
it would not be surprising if the $y$-dependence of $f^{(B\to K^*)}$ and
$g^{(B\to K^*)}$ was consistent with a simple pole in this region.  Between
$y=1$ and $y=3.05$ the form factor $g^{(B\to K^*)}$ falls by roughly a factor
of 3.  In the spacelike region $0 < -Q^2 <1\,{\rm GeV}^2$, over which the pion
electromagnetic form factor falls by a factor of 2.7, its measured
$Q^2$-dependence is consistent with a simple $\rho$
pole~\cite{pion}.\footnote{At higher $-Q^2$, it does appear to be falling
somewhat faster.}  Note also that if $g^{(B\to K^*)}$ and $f^{(B\to K^*)}$ have
pole forms then the $y$-dependence of $g_+^{(B\to K^*)}$ given by
Eq.~(\ref{spinrel2}) does not correspond to a simple pole.

Using Eqs.~(\ref{ffexp}), (\ref{spinrel2}), and (\ref{BDrel}) we obtain
$g_+^{(B\to K^*)}(3.05) = 0.38$.  Then Eq.~(\ref{Rrate}) gives the following
prediction for the $\BKsg$ branching fraction
\begin{equation}\label{prediction}
  {\cal B}(\BKsg) = 4.1 \times 10^{-5} \,.
\end{equation}
To evaluate Eq.~(\ref{Rrate}), we used $\tau_B = 1.6\,$ps, $|C_7| = 0.31$,
$|V_{tb} V_{ts}^*| = 0.04$,  and ${\overline m}_b = 4.2\,$GeV.  This result
compares unexpectedly well with the CLEO measurement ${\cal B}(\BKsg) = (4.2
\pm 0.8 \pm 0.6) \times 10^{-5}$~\cite{CLEO}, and lends support to the validity
of heavy quark symmetry relations between $B$ and $D$ semileptonic form factors
and to the hypothesis that the pole form can be extended beyond $y=1.3$.  Of
course, it is also possible that the agreement between our prediction and data
is a result of a cancellation between large corrections.  Note that the sign of
the form factor $g^{(D\to K^*)}(y)$, which only enters differential
distributions but not the total $D\to K^*$ rate, is very important for the
prediction in Eq.~(\ref{prediction}).  

This set of approximations together with neglecting $SU(3)$ violation in the
form factors $f^{(H\to V)}$ and $g^{(H\to V)}$ also imply that the short
distance contribution to $\Brg$ branching ratio is ${\cal B}(\Brg) = 0.80\,
|V_{td} / V_{ts}|^2 \times {\cal B}(\BKsg)$.

Including perturbative strong interaction corrections, the right-hand-side of
Eq.~(\ref{BDrel}) gets multiplied by $1 + (\alpha_s/\pi) \ln(m_b/m_c)$, but
Eqs.~(\ref{spinrel}) and (\ref{spinrel2}) remain unaffected.  Evaluating
$\alpha_s$ at the scale $\sqrt{m_b m_c}$, this gives a 10\% increase in the
prediction for $g_+^{(B\to K^*)}$ and a 20\% increase in the prediction for the
$\BKsg$ branching ratio in Eq.~(\ref{prediction}).

The factors of $m_D$ and $m_B$ in Eq.~(\ref{BDrel}) are kinematical in origin. 
At $y$ near 1, the validity of Eq.~(\ref{BDrel}) relies partly on the charm
quark being heavy enough that the $B$ and $D$ hadrons have similar
configurations for the light degrees of freedom.  Even though $m_{K^*}/m_D \sim
1/2$, the typical momenta of the ``spectator" light valence quark in the $K^*$
meson is of order $\lqcd$.  Near $y=1$ the corrections to Eq.~(\ref{BDrel})
need not be larger than the order $\lqcd/m_{c,b}$ corrections that occur in
some of the $B\to D^{(*)}$ or $\Lambda_b\to \Lambda_c$ semileptonic decay form
factors.  For example, the $1/m_c$ corrections in the matching of the full QCD
weak current onto the current in the heavy quark effective theory (HQET)
result in the following correction to the form factor $g^{(D\to K^*)}$
\begin{equation}\label{deltag}
\delta g^{(D\to K^*)} = \frac1{4m_c}\, \bigg[ 4\, c^{(D\to K^*)} +
  \bigg(1+{\bar\Lambda\over m_D}\bigg)\, g_+^{(D\to K^*)} +
  \bigg(1-{\bar\Lambda\over m_D}\bigg)\, g_-^{(D\to K^*)} \bigg] \,,
\end{equation}
where $c^{(H\to V)}$ is defined by the HQET matrix element
\begin{equation}
\langle V(p',\epsilon) |\, \bar q\, iD_\mu\, Q\, | H(p) \rangle =
  i\, \varepsilon_{\mu\nu\lambda\sigma}\,  c^{(H\to V)}\, 
  \epsilon^{*\nu}\, (p+p')^\lambda\, (p-p')^\sigma \,.
\end{equation}
The function $c^{(H\to V)}$ is not known, but it could be computed in lattice
QCD.  Neglecting it, and using Eqs.~(\ref{ffexp}) and (\ref{spinrel}) with
$B\to D$, we find that $\delta g^{(D\to K^*)} / g^{(D\to K^*)}$ is about
$\{-0.20,\, -0.13\}$ at $y=\{1,\, 1.3\}$.  It is not surprising that heavy
quark symmetry is useful near $y=1$, but at $y=y_0$ there is no obvious reason
why the relation between $g^{(D\to K^*)}$ and $g^{(B\to K^*)}$ in
Eq.~(\ref{BDrel}) should be valid.  Strictly speaking, our prediction for
$\Gamma(\BKsg)$ does not depend on this assumption.  As long as
Eg.~(\ref{BDrel}) holds for $1<y<1.3$ and the $B$ decay form factors have the
pole form for $y>1.3$, Eq.~(\ref{prediction}) follows.  We do not need to
assume that the $D$ decay form factors also continue to be dominated by the
nearest pole for $y>1.3$ (which is beyond the $\DKsln$ kinematic range). 
Nonetheless, under the assumption that the pole form continues to hold for the
$D$ decay form factors, the order $\lqcd/m_c$ contribution to $\delta g^{(D\to
K^*)} / g^{(D\to K^*)}$ from the last two terms in Eq.~(\ref{deltag}) is not
anomalously large even at $y=y_0$.

If we take Eq.~(\ref{prediction}) as (circumstantial) evidence that heavy quark
symmetry violation in scaling the $g$ and $f$ form factors from $D$ to $B$
decay is small, this has implications for extracting $|V_{ub}|$ from $\Brln$. 
The measurement ${\cal B}(D\to \rho^0 \bar\ell \nu) / {\cal B}(D\to \bar K^{*0}
\bar\ell \nu) = 0.047\pm0.013$ \cite{E791a} suggests that $SU(3)$ symmetry
violation in the $D\to V$ form factors is also small.  Assuming $SU(3)$
symmetry for these form factors, but keeping the explicit $m_V$-dependence in
the matrix element and in the phase space, the measured form factors in
Eq.~(\ref{ffexp}) imply ${\cal B}(D\to \rho^0 \bar\ell \nu) / {\cal B}(D\to
\bar K^{*0} \bar\ell \nu) = 0.044$~\cite{lsw}.\footnote{This prediction would
be $|V_{cd}/V_{cs}|^2/2 \simeq 0.026$ with $m_\rho = m_{K^*}$.  Phase space
enhances $D\to\rho$ compared to $D\to K^*$ to yield the quoted prediction.}

The differential decay rate for semileptonic $B$ decay (neglecting the lepton
mass, and not summing over the lepton type $\ell$) is
\begin{equation}\label{SLrate}
{{\rm d}\Gamma(\Brln)\over{\rm d}y} 
  = {G_F^2\,|V_{ub}|^2\over48\,\pi^3}\, m_B\, m_\rho^2\, S^{(B\to\rho)}(y) \,.
\end{equation}
Here $S^{(H\to V)}(y)$ is the function
\begin{eqnarray}\label{shape}
S^{(H\to V)}(y) &=& \sqrt{y^2-1}\, \bigg[ \Big|f^{(H\to V)}(y)\Big|^2\,
  (2+y^2-6yr+3r^2) \\*
&+& 4{\rm Re} \Big[a_+^{(H\to V)}(y)\, f^{*(H\to V)}(y)\Big]
  m_H^2\, r\, (y-r) (y^2-1) \nonumber\\*
&+& 4\Big|a_+^{(H\to V)}(y)\Big|^2 m_H^4\, r^2 (y^2-1)^2 + 
  8\Big|g^{(H\to V)}(y)\Big|^2 m_H^4\, r^2 (1+r^2-2yr)(y^2-1)\, \bigg] ,
  \nonumber
\end{eqnarray}
with $r=m_V/m_H$.  $S^{(B\to\rho)}(y)$ can be estimated using combinations of
$SU(3)$ flavor symmetry and heavy quark symmetry.  $SU(3)$ symmetry implies
that the $\bar B^0\to\rho^+$ form factors are equal to the $B\to K^*$ form
factors and the $B^-\to\rho^0$ form factors are equal to $1/\sqrt2$ times the
$B\to K^*$ form factors.  Heavy quark symmetry implies the relations in
Eq.~(\ref{BDrel}) and~\cite{IsWi}
\begin{equation}
a_+^{(B\to K^*)}(y) = \frac12 \left({m_D\over m_B}\right)^{1/2} 
  \left[ a_+^{(D\to K^*)}(y) \left(1+{m_D\over m_B}\right) - 
  a_-^{(D\to K^*)}(y) \left(1-{m_D\over m_B}\right) \right] . 
\end{equation}

In the large $m_c$ limit, $(a_+^{(D\to K^*)} + a_-^{(D\to K^*)}) / (a_+^{(D\to
K^*)} - a_-^{(D\to K^*)})$ is of order $\Lambda_{\rm QCD}/m_c$, so we can set
$a_-^{(D\to K^*)} = -a_+^{(D\to K^*)}$, yielding
\begin{equation}\label{a+rel}
a_+^{(B\to K^*)}(y) = \left({m_D\over m_B}\right)^{1/2}\,
  a_+^{(D\to K^*)}(y) \,.
\end{equation}
Eq.~(\ref{a+rel}) may have significant corrections.  In the large $m_c$ limit,
$(g_+^{(D\to K^*)} + g_-^{(D\to K^*)}) / (g_+^{(D\to K^*)} - g_-^{(D\to K^*)})$
is also of order $\Lambda_{\rm QCD}/m_c$.  From Eq.~(\ref{spinrel}) with $B\to
D$ and Eq.~(\ref{ffexp}) we find that $g_-^{(D\to K^*)} = - \lambda g_+^{(D\to
K^*)}$, where $\lambda=\{0.86,\, 1.04\}$ at $y=\{1,\, 1.3\}$.

\begin{figure}[t]
\centerline{\epsfysize=7cm\epsffile{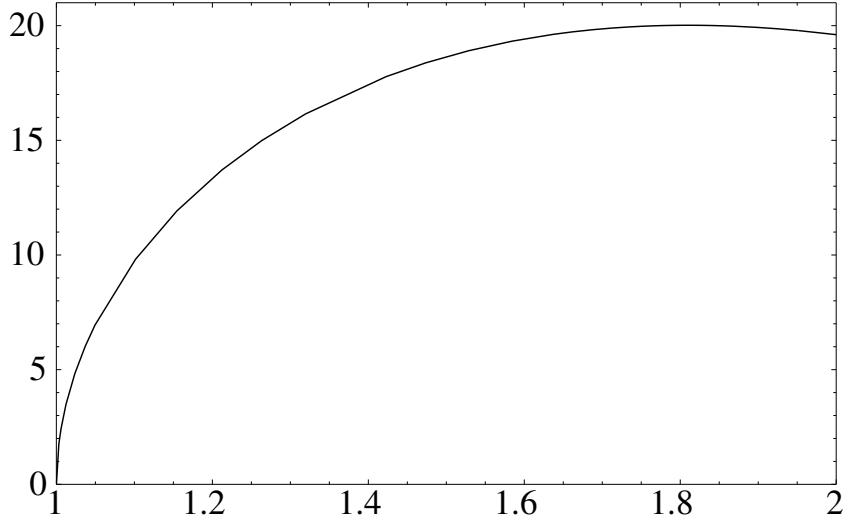}}
\tighten{\caption{$S^{(B\to\rho)}(y)$ defined in Eq.~(\ref{SLrate}) using
the measured $\DKsln$ form factors plus heavy quark and $SU(3)$ symmetry. }}
\end{figure}

Using Eqs.~(\ref{BDrel}) and (\ref{a+rel}), and $SU(3)$ to get the $\bar B^0\to
\rho^+ \ell \bar\nu_\ell$ form factors from those for $\DKsln$ given in
Eq.~(\ref{ffexp}) yields $S^{(B\to\rho)}(y)$ plotted in Fig.~1 in the region
$1<y<2$.  In this region $a_+^{(B\to\rho)}$ and $g^{(B\to\rho)}$ make a modest
contribution to the differential rate.  For $y>2$, $S^{(B\to\rho)}(y)$ is quite
sensitive to the form of $a_+^{(B\to K^*)}$ in Eq.~(\ref{a+rel}) which relies
on setting $a_-^{(D\to K^*)} = -a_+^{(D\to K^*)}$.  An extraction of $|V_{ub}|$
from $\Brln$ data using Fig.~1 in the limited range $1<y<1.3$ is model
independent, with corrections to $|V_{ub}|$ first order in $SU(3)$ and heavy
quark symmetry breaking.  Extrapolation to a larger region increases the
uncertainties both because the sensitivity to setting $a_-^{(D\to K^*)} =
-a_+^{(D\to K^*)}$ increases and because the dependence on the functional form
used for the extrapolation of the form factors increases.  The region $1<y<2$
which contains about half of the phase space has less model dependence than
using the full kinematic region.

In summary, we predicted in Eq.~(\ref{prediction}) the $\BKsg$ branching
fraction in surprising agreement with CLEO data using $b$ quark spin symmetry
at large recoil to relate the tensor and (axial-)vector form factors, using
heavy quark flavor symmetry to relate the $B$ decay form factors to the
measured $\DKsln$ form factors, and extrapolating the semileptonic $B$ decay
form factors to large recoil assuming nearest pole dominance.  Although this
agreement could be accidental, it suggests that heavy quark symmetry can be
used to relate $D$ and $B$ semileptonic form factors and that $f^{(B\to K^*)}$
and $g^{(B\to K^*)}$ can be extrapolated to $y>1.3$ using the pole form.  This
is encouraging for the extraction of $|V_{ub}|$ from $\Brln$ using Fig.~1.  If
experimental data on the $\Drln$ and $\BKsll$ differential decay rates become
available, then a model independent determination of $|V_{ub}|$ can be made
with corrections only of order $m_s/m_{c,b}$ (rather than $m_s/\lqcd$ and
$\lqcd/m_{c,b}$)~\cite{lwold,lsw,ben}.

\acknowledgements

We thank Jeff Richman for a conversation that led to this paper, and Adam Falk
for useful remarks.
M.B.W.\ was supported in part by the U.S.\ Dept.\ of Energy under Grant no.\
DE-FG03-92-ER~40701.  Fermilab is operated by Universities Research
Association, Inc., under DOE contract DE-AC02-76CH03000.

{\tighten

} %end tighten 

\end{document}